\documentclass[prd, preprintnumbers,amsmath,amssymb, nofootinbib]{revtex4}

\usepackage{graphicx}
\usepackage{amsmath}
\usepackage{epsfig, epstopdf}
\usepackage{amsmath,amsfonts,amssymb}

\newcommand{\nc}{\newcommand}

\usepackage{color}
\nc{\ba}{\begin{eqnarray}}
\nc{\ea}{\end{eqnarray}}

\newcommand{\calR}{{\cal{R}}}
\newcommand{\calP}{{\cal{P}}}

\nc{\x}{{\bf{x}}}
\nc{\bfk}{{\bf{k}}}

\def\be{\begin{equation}}
\def\ee{\end{equation}}
\def\bea{\begin{eqnarray}}
\def\eea{\end{eqnarray}}

\def\bearst{\begin{eqnarray*}}
\def\eearst{\end{eqnarray*}}

\begin{document}

\title{Fingerprints of Anomalous Primordial Universe \\ on the Abundance of Large Scale Structures}

\author{Shant Baghram$^{1,2}$}
\email{baghram-AT-sharif.edu}
\author{Ali Akbar Abolhasani$^{1,3}$}
\email{abolhasani-AT-ipm.ir}
\author{Hassan Firouzjahi$^{2}$}
\email{firouz-AT-mail.ipm.ir}
\author{Mohammad Hossein Namjoo$^{4,2}$}
\email{MohammadHossein.Namjoo-AT-utdallas.edu}

\affiliation{$^{1}$Department of Physics, Sharif University of
Technology, P.~O.~Box 11155-9161, Tehran, Iran}

\affiliation{$^{2}$School of Astronomy, Institute for Research in
Fundamental Sciences (IPM),
P.~O.~Box 19395-5531,
Tehran, Iran}

\affiliation{$^{3}$School of Physics, Institute for Research in
Fundamental Sciences (IPM),
P.~O.~Box 19395-5531,
Tehran, Iran}

\affiliation{$^{4}$
Department of Physics, The University of Texas at Dallas, Richardson, TX 75083, USA}

\vskip 1cm

\begin{abstract}

We study the predictions of anomalous inflationary models  on the abundance of structures
in large scale structure observations. The anomalous features encoded in primordial curvature perturbation power spectrum  are (a): localized feature in momentum space, (b): hemispherical asymmetry   and (c): statistical anisotropies.  We present a model-independent expression relating the
number density of structures to the changes in  the matter density variance. Models with localized feature can alleviate the tension between observations and numerical simulations of cold dark matter structures on galactic scales as a possible solution to the  missing satellite problem. In models with hemispherical asymmetry we show that  the  abundance of structures becomes asymmetric depending on the direction of observation to sky. In addition, we study the effects of scale-dependent dipole amplitude on  the abundance of  structures.  Using the quasars data and adopting the power-law scaling $k^{n_A-1}$ for the amplitude of dipole we find the upper bound $n_A<0.6$ for the spectral index of the dipole asymmetry.  In all cases there is a critical mass scale $M_c$  in which for $M<M_c (M> M_c)$ the enhancement in variance induced from anomalous feature decreases (increases)  the abundance  of dark matter structures in Universe.

\end{abstract}

\maketitle


\section{Introduction}
\label{Sec-I}
The standard model of cosmology, $\Lambda$CDM, explains anisotropies on  Cosmic Microwave Background (CMB) radiation \cite{Ade:2013zuv} and the Large Scale Structure (LSS) observations with good accuracies \cite{Tegmark:2003uf}. This model is based on the assumption of cold dark matter as the skeleton of the structures in the Universe  \cite{Frenk:2012ph} and the cosmological constant as the source of late time acceleration while the initial conditions are set by  the inflationary paradigm \cite{Guth:1980zm,Sato:1980yn,Linde:1981mu,Albrecht:1982wi}. Inflation generates nearly scale-invariant, nearly Gaussian and nearly adiabatic perturbations
which are imprinted  on the temperature map of CMB and also provides the initial distribution for the seeds of structures in the early Universe \cite{Komatsu:2010fb}. Accordingly, one of the triumphs of modern cosmology is  the compatibility of the dark matter tracers distribution with the statistics of CMB temperature anisotropy \cite{Tegmark:2003ud}.

Thanks to vast cosmological data available during last decade, models of early universe are well-constrained by cosmological observations. In particular, the high precision maps of CMB have been studied extensively to put constraints on inflationary models \cite{Komatsu:2010fb, Hinshaw:2012aka, Ade:2013uln}. Any deviation from the Gaussian, adiabatic, isotropic and scale-invariant initial conditions will open distinct fingerprints on CMB map as tools
to study the physics of early universe such as inflation  \cite{Komatsu:2010hc, Chen:2010xka, Wang:2013zva}.   On the other hand, the LSS observables, by transferring the initial condition of perturbations to the late time observables,  are potentially very useful tools to test models of early universe in sub-CMB scales  \cite{Komatsu:2009kd}. The abundance of the structures in the Universe \cite{Mana:2013qba,Hazra:2012vs}, the scale-dependence of bias parameter between dark matter halos as the host of galaxies and the dark matter perturbations \cite{Dalal:2007cu} and higher order statistics of dark matter tracers (e.g. the bispectrum of galaxies) \cite{Sefusatti:2007ih} are all used to probe the deviations from the standard initial conditions.

There are indications for anomalous features in CMB map such as power asymmetry between the two CMB hemispheres \cite{Ade:2013nlj},  power deficit in low multipoles
\cite{Ade:2013uln} and the coincidence of the quadrupole and octupole directions \cite{Ade:2013nlj}. These observations trigger the motivation to study the predictions of inflationary models with anomalous features  in LSS observables. We note that the CMB maps cover the first few e-folds of the inflationary period, say the first 5-7 e-folds, while the remaining 50 or so e-folds are below the CMB scales. It is likely that the primordial power spectrum may have non-trivial features not only on CMB scales but also on scales which are beyond the probe of CMB but are accessible to LSS observations.   In this work we investigate the framework under which one can study the probable observational fingerprints of anomalous inflationary power spectrum on LSS,  specially in the abundance of the structure.  Having this said, there are open questions in LSS, specially on galactic scale, such as  the ``missing satellite" problem \cite{Bullock:2010uy} and the ``too big to fail" problem \cite{BoylanKolchin:2011de} which  may be considered as hints for anomalies/features in primordial curvature perturbation power spectrum on galactic scales (for a recent work addressing this idea see \cite{Garrison-Kimmel:2014kia}). It is worth to mention that the theoretical predictions of standard model for matter power spectrum are in reasonable agreements with the observational data in the range of wavenumber (scales)  $0.01 hMpc^{-1}< k < 1 hMpc^{-1}$. However, the amplitude of
small scale dark matter power spectrum  is not strongly constrained by LSS observables.
Accordingly, there are freedoms on the amplitude of the power on galactic scales which in turn
can be used as an explanation for the missing satellite problem. Therefore  the anomalous inflationary models, by modifying the power on small scales, can be considered as a possible solutions to the tension between observations and theory on galactic scales.

The rest  of the paper is organized as follows. In Section \ref{Sec-II} first we review  the procedure for obtaining the number density of structures based on Excursion Set Theory \cite{Bond:1990iw} and then we study the effects of the primordial feature on the number density of structures in a model-independent way. In Section \ref{theory} we present the theoretical motivations for generating primordial features studied in this paper.
In Section \ref{Sec-III}, we study the predictions of each anomalous inflationary model on the abundance of the structures. The reader who is not interested in theoretical realization of the features can skip Section \ref{theory} and go directly to Section \ref{Sec-III}.
The conclusion and future prospects are presented in Section \ref{Sec-IV}.

\section{ The number density of structures}
\label{Sec-II}

The mass function of the dark matter halo structures  can be found from the Excursion Set Theory (EST) as a barrier crossing problem of density perturbation in different mass scales (window functions) \cite{Bond:1990iw,Zentner:2006vw}. The number density $n(M,z)$ of virialized objects with the mass  between $M$ and $M+dM$ is  \cite{Mo:1995cs}
\be \label{Eq:massfunc}
n(M,z)=-2\frac{\bar{\rho}}{M^2}f(\nu)\frac{d\ln\sigma(M,z)}{d\ln M}
\ee
where $\bar{\rho}$ is the mean matter density  and $f(\nu)$ is the universality function of halo distribution which is a function of height parameter $\nu\equiv \delta_c/\sigma(M,z)$. Here  $\delta_c\simeq 1.68$ is  the critical density (in spherical collapse) \cite{Gunn:1972sv} and
$\sigma(M,z)$ is the variance of the structures with mass $M$ at the specific redshift $z$.

The variance is defined as
\be\label{Eq:sigma}
 \sigma^2(M,z)=\int\frac{d^3 k}{(2\pi)^3} P_L(k,z) W^2(kR)=\int\frac{dk}{2\pi^2}k^2P_L(k,z)W^2(kR)
\ee
where $P_L(k)$ is the linear matter power spectrum and the second equality in Eq. (\ref{Eq:sigma}) is obtained with the assumption of isotropic density perturbations. In addition, $W(x=kR)$ is the window function, which for the top-hat filter in real space, has the following form in momentum space
\be
W(x)=3\frac{\sin(x)-x\cos(x)}{x^3} \, ,
\ee
in which modes with $k> R^{-1}$ are smoothed out.

The variance  obtained for each mass scale is related to the radius of window function via
the relation  $M=4\pi \bar{\rho}R^3/3$. An important point to mention is the universality of mass function, i.e. it  is only a function of the height parameter $\nu$. The universality means that on each mass range we have the same dependence on the height parameter. The height parameter $\nu$ encodes the information about the local threshold of collapse as well as the  variance of  matter perturbations which the latter depends on the cosmological model. The Press-Schechter mass function, as one of the first proposals for the universality function, is defined as \cite{Press:1973iz}
\be
f(\nu)=\frac{\nu}{\sqrt{2\pi}}\,  e^{-\nu^2/2} \, ,
\ee
assuming that the primordial seeds of the structures have a Gaussian distribution. The N-body simulations have investigated the validity of the Press-Schechter conjecture and have found that it works with $10\%$ accuracy. The Sheth-Tormen mass function, incorporating the idea of ellipsoidal collapse \cite{Sheth:1999su}, has the universality function $f(\nu)$ given by  \cite{Sheth:1999mn}
\be
f(\nu)= C \, \sqrt{\frac{a}{2\pi}}\nu\left[1+({a\nu ^2})^{-p}\right]e^{-a\nu^2/2} \, ,
\ee
where $a\simeq 0.7$ and $p\simeq 0.3$ are constants which are determined from N-body simulations of LSS and $C$ is determined from the  requirement that $-2 \int_0^\infty  f(\nu)/\nu =1$.
In the limit $C=\frac{1}{2}$, $p=0$ and $a=1$, we recover the Press-Schechter proposal.
The discrepancy between Press-Schechter and the results obtained from the N-body simulation indicates the complexity of non-linear structure formation. In addition, it may indicate that the assumptions such as the spherical collapse or the Markovianity of the random walks of  matter density perturbations in EST framework are only approximation to the real process of collapse. There is a vast literature on the prediction of dark matter clustering in EST framework with deviations from Gaussianity, Markovianity and spherical collapse \cite{Maggiore:2009rv,Maggiore:2009rw, Maggiore:2009rx,DeSimone:2010mu,D'Amico:2010ta}. In this work, in order to simplify the analysis and also in order to get some intuitions on the effects of anomalous features on LSS abundances,  we consider only the Press-Schechter and Sheth-Tormen universality functions.

In order to study  the effects of  feature in primordial curvature perturbations on the abundance of the structures we should relate the late-time linear matter power spectrum to the initial curvature perturbations power spectrum.  The Poisson equation relates the matter density contrast in Fourier space, $\delta(k,z)$, to the gravitational potential   (Bardeen potential) $\Phi(k,z)$ via
\be
\delta_m (k,z) =\frac{2}{3} \frac{k^2\Phi(k,z)}{\Omega_m^0 H_0^2(1+z)} \, ,
\ee
where $\Omega_m^0$ is the matter density parameter at the present time (hereafter we omit the superscript $0$ in density parameter) and $H_0$ is the present value of Hubble parameter.
Now we can relate the gravitational  potential $\Phi(k,z)$ to the primordial curvature perturbation ${\cal{R}}_k$ by the transfer function $T(k)$ and the growth function $D(z)$ (which are separable functions in $\Lambda$CDM model) via
\be
\Phi(k,z)=\frac{3}{5}T(k)D(z)(1+z){\cal{R}}_k \, .
\ee
Note that we set the primordial (initial) Bardeen potential in radiation dominated era as $\Phi_{ini}(k,z_{ini})=2{\cal{R}}_k/3$.
Consequently, the matter power-spectrum is related to the primordial  curvature perturbation via
\be
P_L(k,z)=\frac{8\pi^2}{25}k\frac{T^2(k)D^2(z)}{\Omega^2_m H_0^4}{\cal{P}}_{\cal{R}}(k) \, ,
\ee
where ${\cal{P}}_{\cal{R}}\equiv \frac{1}{2\pi^2}k^3P_{\cal{R}}(k)$ is the dimensionless power spectrum.  Usually the curvature perturbations power spectrum is  parameterized as
\be
{\cal{P}}_{\cal{R}}={\cal A} (\frac{k}{k_p})^{n_s-1} \, ,
\ee
where $ {\cal A}$ is the amplitude of the fluctuations at pivot wavenumber $k_p=0.05 Mpc^{-1}$ (this value is from Planck collaboration) and $n_s$ is the spectral index which shows the scale-dependence of power spectrum. The constraints from the  Planck collaboration are  $10^9\times {\cal A}=2.23\pm 0.16$ and $n_s=0.9616\pm 0.0094$ at $1\sigma$ \cite{Ade:2013zuv}.

The question we are interested to answer is  {\it{if we put a feature in primordial curvature perturbations power spectrum, how does this affect the abundance of the structures in LSS?}}  With the inclusion of feature in power spectrum, we schematically  have
\ba \label{Eq:power-mod}
\calP_\calR ({\bf{k}}) = \overline{\calP_\calR} ({\bf{k}}) + \delta \calP_\calR(\bfk) \, ,
\ea
in which  $ \overline {\calP_\calR}$ is the standard isotropic and nearly scale-invariant  primordial curvature perturbation power spectrum and $\delta \calP_\calR(\bfk)$ is a scale-dependent feature added to the curvature power. Consequently, the variance of matter perturbation from Eq.(\ref{Eq:power-mod}) is modified as follows
\be
\sigma(M,z; A_*)=\bar{\sigma}(M,z)+\delta\sigma(M,z;A_*) \, ,
\ee
where $\bar{\sigma}(M)$ is the variance in $\Lambda CDM$ model obtained from $\overline {\calP_\calR}$ and $\delta\sigma$ is its modification induced from the correction in primordial power spectrum. In this notation, $A_*$ is a symbolic parameter which collectively represents  the set of the parameters which define the feature (e.g. the  amplitude  of the feature or the location of feature in momentum space).

The leading order correction in the variance is given by
\ba \label{Eq:deltasigma0}
\delta \sigma (M,z;A_*) = \dfrac{2}{25 \bar \sigma} \dfrac{D^2(z)}{\Omega_m^2 H_0^4}
\int  \dfrac{k^3 dk}{4 \pi} d\Omega_k W^2(kR) T^2(k) \delta \calP_\calR(\bfk;A_*) \, ,
\ea
where $d\Omega_k$ is the solid angle  of the spherical coordinates in  the Fourier space.  In the case in which the correction to the power spectrum is isotropic we have
\ba \label{Eq:variance-change}
\delta \sigma (M,z;A_*) = \dfrac{2}{25 \bar \sigma} \dfrac{D^2(z)}{\Omega_m^2 H_0^4}
\int  k^3 dk W^2(kR) T^2(k) \delta \calP_\calR(k;A_*)  \quad \quad
\mathrm{( isotropic ~\,  feature )}  \, .
\ea
The next step is to translate the effects of the corrections in variance to  the number density of structures. For  this purpose, we use the fact that the variance $\bar \sigma$ in $\Lambda$CDM model has a power-law relation with the radius (Mass) of the enclosed over-dense region as follows \cite{Viana:1995yv,White:1996pz,Liddle2000}
\be \label{sigma8}
\bar{\sigma}_R(z)\simeq \sigma_*(z)(\frac{R}{R_*h^{-1}Mpc})^{-\gamma}
\ee
where $\sigma_*$ is the normalization factor  and $\gamma$ is the fitting parameter. For the mass range $M=(10^9-10^{15}) M_{\odot}$, $R_*=8h^{-1}Mpc$, accordingly the normalization parameter is $\sigma_*=\sigma_8\equiv\sigma(R=8h^{-1}Mpc)$ and the parameter $\gamma$  is approximately given by $\gamma=(0.3\Gamma + 0.2) \left[2.92+\log(\frac{R}{8h^{-1}Mpc}) \right]$ in which to leading order $\Gamma \simeq 0.25$ \cite{White:1996pz}. For the mass range  $M<10^9M_{\odot}$, $R_*=0.1h^{-1}Mpc$, accordingly the normalization parameter is $\sigma_*=0.05\times\sigma_{0.1}\equiv0.05 \times\sigma(R=0.1h^{-1}Mpc)$ and the parameter $\gamma\simeq0.1 + {\cal{O}}(\log[\frac{R}{0.1h^{-1}Mpc}])$. Neglecting the logarithmic dependence of the variance to the mass ( this approximation introduces an error of less than $10\%$), in these two intervals the mass function of the structures becomes
\be \label{eq:fwn}
{\bar{n}(M,z)} \simeq  \frac{2}{3}\gamma\frac{\bar{\rho}}{M^2}f(\nu(z)) \, ,
\ee
where $\gamma$ is almost constant and depends on the mass interval we are interested. In Eq.(\ref{eq:fwn})  $\bar{n}$ is the number density per mass of  structures with the standard initial conditions, i.e. in the absence of the feature.

Now we define a new parameter, $\delta _n$, as the fractional excess of the number density of the structures in the mass range $M$, $M+dM$ as follows
\be \label{Eq:deltan}
\delta _ n (M,z; A_*)= \frac{n(M,z; A_*)-\bar{n}(M,z)}{\bar{n}(M,z)}
\ee
where $n(M,z; A_*)$ is the mass function of the structures in the presence of the feature   in primordial curvature power spectrum collectively  represented by the parameter $A_*$.
We also consider the fact that the number density of the structures can be calculated at any redshift.
Therefore,  the change in number density depends on the redshift through the growth function
$D(z)$ which appears  in the variance of the matter perturbation.

Now, assuming the universality of $\bar{n}(M,z)$ which is encoded in $f(\nu)$, $\delta_n$ at any mass scale is obtained  to be
\be \label{Eq:deltan1}
\delta _n (M,z; A_*)=\frac{f'(\nu (M,z))}{f(\nu (M,z))}\frac{\partial\nu}{\partial\bar{\sigma}}\delta\sigma  (M,z;A_*) \, ,
\ee
where a prime represents the  derivative with respect to the height parameter $\nu$. Note that
Eq. (\ref{Eq:deltan1}) is obtained from the Taylor expansion of Eq. (\ref{Eq:deltan}) with respect to change in variance induced from $A_*$ at a fixed mass. Using the definition of height parameter, Eq. (\ref{Eq:deltan1}) reduces to
 \be \label{deltan1}
 \delta _n(M,z; A_*)= -\nu\frac{f'(\nu (M,z))}{f(\nu (M,z))}\frac{\delta\sigma  (M,z;A_*)}{\bar{\sigma} (M,z)} \, .
 \ee
From this equation  we see that the change in number density, $\delta_n$, is proportional  to the change in the variance $\delta\sigma/\bar{\sigma}$ induced by the primordial feature. We note that the proportionality factor is given by the logarithmic derivative of the universal   mass function. Consequently, an excess or deficit in variance does not have a uniform effect on all mass scales;  its effect is non-trivial which  depends on the mass scale under consideration.

For the  Press-Schechter universality function we have
\be  \label{nps}
 \delta _n |_{PS}(M,z;A_*)= \left(\nu^2(M,z)-1 \right)\frac{\delta\sigma(M,z;A_*)}{\bar{\sigma}(M,z)} \, ,
\ee
where $ \delta_n|_{PS}(M,z;A_*)$ is the change in number density at a specific mass $M$. This change is a function of the mass through the height parameter.

Now consider the critical height $\nu_c(M_c(z)) =1$ with the critical  mass $M_c$ which depends on redshift.   The interesting point  is that for the structures with the mass range $M(z)>M_c(z)$  in which $\nu(M>M_c)>\nu_c=1$,  the number density contrast has an excess for $\delta\sigma >0$ while opposite conclusion is reached for $M(z)<M_c(z)$
in which $\nu(M<M_c)<1$.  Note that the sign flip for $\delta_n$ at the critical height parameter $\nu_c$ is a result of the Gaussian probability distribution function (PDF) which is assumed in the Press-Schechter formalism. For a Gaussian PDF with a fixed value of the density contrast $\delta$ the probability is maximum if the variance satisfies the relation $\sigma = \delta$. If one fixes the density contrast to the critical value $\delta = \delta_c$,  which is the definition of critical height parameter $\nu_c=1$, then the maximum probability occurs at $\sigma =\delta_c $, or equivalently at $\nu=1$. That is why $\delta_n$ for the collapsed objects as a function of the variance flips the sign at $\nu=1$.  A crucial point to discuss is that for the mass scales $M>M_c$ where $\nu>\nu_c$, we are in quasi-linear regime where the structures are on the verge of formation and the density contrast  is almost near the critical density $\delta_c$ (the uncertainty comes from the fact that the condition of spherical collapse is $\delta>\delta_c$ which is not exactly equal to $\nu<1$). Accordingly, our result is more reliable for masses and redshifts in which $\nu \leq 1$ while it is not applicable for $\nu\gg1$.

Similarly, we can find the number density contrast for the Sheth-Tormen mass function as
\be\label{Eq:deltanST}
 \delta _n |_{ST}(M,z;A_*)= \left[ a\nu^2(M,z)-1+\frac{2ap\nu^2(M,z)(a\nu^2(M,z))^{-p-1}}{[1+(a\nu^2(M,z))^{-p}]} \right] \frac{\delta\sigma(M,z,A^*)}{\bar{\sigma(M,z)}} \, .
\ee
In the case  $p=0$ and $a=1$ the Sheth-Tormen number density contrast reduces to that of  the Press-Schechter obtained in Eq.(\ref{nps}). Interestingly, the critical height in which the big bracket in Eq.(\ref{Eq:deltanST})  vanishes is more complicated than the Press-Schechter function in which one simply has $\nu_c=1$.  This indicates that in a more realistic model of collapse the change in the statistics of dark matter halos is a non-trivial function of the  mass scale $M_c$ which in itself is a redshift dependent quantity.

We remark that in this work we assume that  introducing a feature in primordial power spectrum does not invalidate the Press-Schechter procedure of structure formation. This means that the processes of barrier crossing, the magnitude of $\delta_c$ obtained from spherical collapse and the
environment-independence of standard EST do not change dramatically in the models we study in next section and the corrections are sub-leading. This is because the mechanism of collapse occurs in late time Universe and it is almost scale-free. The  inflationary models with feature mainly change the initial conditions of perturbations which to first order modifies the  variance as we have discussed here.

With these model-independent analysis for $\delta \sigma/\bar\sigma$ and $\delta_n$, in Section \ref{Sec-III} we  calculate  these quantities  for various features in curvature perturbation power spectrum presented in Section \ref{theory}.

\section{ Inflationary Models  with Anomalous Features}
\label{theory}

In this Section we present three different kinds of anomalous features in inflationary power spectrum which will be used for the abundance of structures in next Section. The reader who is not interested in mechanism of generating these  features in inflationary model building can directly jump to Section \ref{Sec-III}. Our discussions here will be brief, we only provide the motivations and the overviews of  mechanisms  generating the corresponding features. For detail studies, we refer the reader to the extensive literature on each topic.

\subsection{ Localized Features in Power Spectrum}
\label{local-sec}

First we consider models of inflation in which there is a sharp jump in
curvature perturbations power spectrum in Fourier space. Mathematically, we parameterize this
localized feature in power spectrum by
\be
\label{Eq-delta}
{\cal{P}}_{{\cal{R}}}=\overline{{\cal{P}}}_{{\cal{R}}} \left[1+A_*\delta_D(k-k_*) \right ]
\ee
in which $\delta_D(k-k_*)$ is the Dirac delta function,  $\overline{\cal{P}}_{{\cal{R}}}$ is the usual nearly scale-invariant power spectrum while the
amplitude of feature is represented by $A_*$ in units of $h Mpc^{-1}$.
In order for this localized feature to be relevant for LSS observations, we require $k_* \ge 1h Mpc^{-1} $. We note that $A_*$ can be either positive or negative, corresponding respectively to an increase or  a decrease in  the primordial power spectrum.

The idea of generating local features have been extensively studied in the literature. Here we follow the mechanism employed in  \cite{Abolhasani:2012px}.  The model contains two scalar fields $\phi$ and  $\chi$ in which the light field $\phi$ is the inflaton field and the heavy field $\chi$ is the waterfall field. The potential is as in hybrid inflation \cite{Linde:1993cn, Copeland:1994vg} given by
\ba
\label{potential}
V=  \frac{m^2}{2} \phi^2
 + \frac{\lambda}{4} \left( \chi^2 - \frac{M^2}{\lambda} \right)^2
+ \frac{g^2}{2} \phi^2 \chi^2  \, .
\ea
However, unlike models of hybrid inflation, in this scenario inflation  mainly  proceeds as in
chaotic model with potential $V \simeq m_{eff}^2 \phi^2/2$
with some effective mass $m_{eff}$ which undergoes a
small but abrupt change when $\phi$ reaches the critical value $\phi=\phi_c\equiv M/g$.  The waterfall field $\chi$ is introduced to induce the change in mass. In this picture inflation has three stages as follows.

As in chaotic models inflation starts  with a large super-Planckian value,  $\phi=\phi_i \gg M_P$ in which $M_P$ is the reduced Planck mass,  so one can obtain about 60 $e$-folds of inflation to solve the horizon and the flatness problems. As in hybrid inflation the waterfall field is very heavy during the first stage of  inflation so it is stuck to its instantaneous minimum $\chi=0$.  However,  once the inflaton field has reached  the critical value
$\phi=\phi_c$, the waterfall field becomes tachyonic and it rapidly rolls towards  its global minimum  $\chi_{min}^2= M^2/\lambda$. This represents the second stage of inflation which may be taken to be very short for our purpose.  The final stage of inflation, corresponding to  $\phi>\phi_c$, follows  as in chaotic inflation but with the effective mass of the inflaton, $m_+$, given by
\ba
m_+^2 = m^2 + g^2 \langle \chi^2 \rangle = m^2 \left( 1+ \Omega \right) \, ,
\ea
in which the constant $\Omega$ is defined via $\Omega\equiv \frac{g^2 M^2}{\lambda m^2}$.

We assume the change in inflaton mass is small, corresponding to  $\Omega \ll 1$, so the slow-roll conditions are satisfied throughout inflation.  In order to bring this local feature into the LSS window
we assume that the short waterfall stage begins when the LSS modes $k_*$ leaves the horizon. Typically, this corresponds to around 50 $e$-folds or so before the end of inflation.

The details of correction in power spectrum and bispectrum have been studied in \cite{Abolhasani:2012px}. The correction in power spectrum from waterfall $\calP_\calR^{wf}$ compared to the original power spectrum coming from the inflaton $\calP_\calR^{\phi}$ is obtained to be
\ba
\label{powers-ratio}
\frac{{\cal P}_\calR^{wf}(k_{*})}{{\cal P}_\calR^{\phi}}
 \simeq 10^{3} \Omega^2 \left(\frac{\epsilon}{10^{-2}}\right)^4
\left(\frac{\epsilon_\chi}{10}\right)^{-10/3} \, ,
\ea
in which $\epsilon$ is the slow-roll parameter and $\epsilon_\chi$ is a parameter defined in \cite{Abolhasani:2012px} measuring the sharpness of the phase transition. For a very sharp waterfall phase transition, we require $\epsilon_\chi \gg 1$, say $\epsilon_\chi =10$.  With $\Omega \sim 0.1$ the above ratio can be as big as 10.  In principle by taking $\epsilon_\chi$ very big, one can make the duration of the waterfall phase transition (the intermediate second stage of inflation) as short as one desires. In this limit, we approach the  localized feature presented in Eq. (\ref{Eq-delta}).

\subsection{Hemispherical Asymmetry in Power Spectrum}
\label{dipole-sec}

There are indications of power asymmetry in CMB map as observed by Planck satellite \cite{Ade:2013nlj}, for earlier reports of hemispherical asymmetry in WMAP data see \cite{Eriksen:2007pc, Hoftuft:2009rq, Hansen:2008ym}. The observations indicate that the power spectrum in northern hemisphere is different than the the power spectrum in southern hemisphere. Using the following  dipole modulation on  the  CMB temperature fluctuations  \cite{Gordon:2006ag}
\be
\label{dipole-power0}
\Delta T (\hat{\bf n})=\overline{ \Delta T (\hat{\bf n}) }\left[1+A_d\cos(\theta_{\hat{\bf n}.\hat{\bf p}}) \right] \, ,
\ee
in which $\overline{ \Delta T (\hat{\bf n}) }$ is the statistically isotropic temperature fluctuations,
$A_d$ is the amplitude of the dipole asymmetry, $ \hat{\bf p}$ is the preferred direction in the sky, $\hat {\bf n}$ is the direction of the look to sky and $\theta_{\hat{\bf n}.\hat{\bf p}}$ is the angle between $ \hat{\bf p}$ and $ \hat{\bf n}$,  the  Planck collaboration has found $A_d=0.07$ and $\hat{\bf p}=(l=227,b=-27)$ in galactic coordinates \cite{Ade:2013nlj}, see also \cite{Akrami:2014eta, Aslanyan:2014mqa,Notari:2013iva}.   Whether this detection is statistically significant is under debate and we shall wait for the upcoming Planck data release  to see if the significance of the detection will improve.

The best known method to generate a hemispherical asymmetry is the idea of long mode modulation \cite{Erickcek:2008sm, Dai:2013kfa}. In this picture, a long super-horizon mode with the wavelength $1/k_L$ modulates the curvature perturbations at the desired scales (either the CMB or the LSS scales). An observer probing a cosmological distance $x$ much smaller than the long mode ($ k_L x \ll 1)$ can not probe the wave-like nature of the long mode. For this observer, the long mode only modifies the background quantities such as $\dot \phi$ for the inflaton field or modulates the surface of end of inflation. This in turn induces a directional-dependence (a gradient) in power spectrum taking the form
\be
\label{dipole-power}
{\cal{P}}_{{\cal{R}}}=\overline{{\cal{P}}}_{{\cal{R}}}\left[1+2 A_d\cos(\theta_{\hat{\bf n}.\hat{\bf p}}) \right] \, .
\ee
Note that the additional factor 2 in the above formula compared to Eq. (\ref{dipole-power0}) is because  $\calP_\calR \sim \langle \Delta T (\hat{\bf n})^2  \rangle $.

For this picture to work, there should exist a coupling between the long mode $k_L$ and the smaller modes i.e. the CMB-scale or the LSS-scale modes. This rings the bell for the relation between the amplitude of dipole asymmetry and  the `` squeezed-limit'' non-Gaussianity \cite{Namjoo:2013fka, Abolhasani:2013vaa, Firouzjahi:2014mwa, Namjoo:2014nra}, however see \cite{Lyth:2014mga} for
a different view.  Imposing the quadrupole constraints one finds \cite{Lyth} $A_d \lesssim 0.01 f_{NL}^{1/2}$.

With these discussions in mind, in order to get large enough dipole asymmetry one has to start with a model with large enough squeezed-limit non-Gaussianity. As is well-known from Maldacena's analysis \cite{Maldacena:2002vr}, many single field models of inflation fail to do this job. Indeed, Maldacena's non-Gaussianity  consistency condition \cite{Maldacena:2002vr} implies that $f_{NL} \sim n_s -1 \sim 0$ so one can not generate observable dipole asymmetry from the long mode modulation in conventional single field models of inflation. This argument also implies that one has to look for non-trivial models of inflation in which the Maldacena's consistency condition is violated such as in curvaton models or in  multiple fields models of inflation.

The observations on sub-CMB scales indicate that dipole asymmetry becomes less significant
 for $\ell >600$  \cite{Hirata:2009ar, Flender:2013jja}.  This indicates a non-trivial scale-dependent dipole asymmetry, i.e. $A_d = A_d (k)$ which should be considered in any theoretical  modeling of dipole asymmetry. We take into account  the upper bound imposed from  \cite{Hirata:2009ar} when consideration the effects of dipole asymmetry on LSS in next section.

\subsection{Statistical Anisotropies}
\label{stat-sec}

Homogeneity and isotropy are two pillars of cosmological principles yielding the standard FRW cosmology. There were indications of statistical anisotropies in WMAP  data \cite{Bennett:2010jb}, but the follow up analysis demonstrated that the systematic errors can accommodate the apparent source of quadrupole  asymmetry \cite{wmapaniso-crit}.  Having this said, the possibilities of violating statistical isotropy is intriguing both
observationally and also from the theoretical point of view. Indeed, it is likely that there is no statistical anisotropy at CMB scales but there may exist statistical anisotropy at LSS scales.

One interesting mechanism to generate  statistical anisotropies are based on the dynamics of gauge fields. The presence of a background gauge field breaks the three-dimensional rotation invariance and the space-time takes the form of Bianchi I metric. Specifically, assuming that the gauge field $A_\mu(t)$ is turned on along the $x$-direction the metric  line element is given by
\ba
ds^2 = -dt^2  + a(t)^2 dx^2 + b(t)^2 ( dy^2 + dz^2) \, ,
\ea
in which $a(t)$ and $b(t)$ are two scale factors with the corresponding Hubble expansion rates
$H_a = \dot a/a$ and $H_b= \dot b/b$.

A vector field in an expanding background suffers from the conformal invariance so its background value dilutes exponentially during inflation. In order to break the conformal invariance and prevent the background gauge field to decay exponentially, one can couple the gauge field non-trivially to inflaton field.  An interesting realization of anisotropic inflation is proposed in \cite{Watanabe:2009ct}
based on the following Lagrangian
\ba
\label{action}
S= \int d^4 x \sqrt{-g} \left [ \frac{M_P^2}{2}
R -  \frac{1}{2} \partial_\mu \phi\partial^\mu \phi - \frac{f^2(\phi)}{4}F_{\mu \nu} F^{\mu \nu}  - V(\phi) \right] \, ,
\ea
where $M_P$ is the reduced Planck mass,  $\phi$ is the inflaton field and $F_{\mu\nu} = \partial_\mu A_\nu - \partial_\nu A_\mu $ is the gauge field strength.  It is shown in \cite{Watanabe:2009ct} that  with a suitably chosen gauge kinetic coupling $f(\phi)$  the inflationary system  reaches an attractor solution in which the gauge field energy density approaches a small but observationally detectable fraction of the total energy density. In is shown that  the anisotropies produced are at the order of slow-roll parameters.

The anisotropic power spectrum in this model has the quadrupole asymmetry given by
\ba
\label{quad-P}
\calP_\calR(k)  = \overline{\calP_\calR } (k) \,  \left( 1+ g_* \cos^2 \theta_{{\bf \hat p} \cdot {\bf \hat k}}
\right)
\ea
in which ${\bf \hat p}$ is the anisotropic direction (the $x$-direction in our example) and
${\bf \hat k}$ is the momentum direction in Fourier space so $\theta_{{\bf \hat p} \cdot {\bf \hat k}}$
represent the angles between these two directions.
With this definition, $g_*$ is the amplitude of quadrupole asymmetry. For the model of anisotropic inflation introduced in \cite{Watanabe:2009ct} one can show that
$g_* \propto N^2$ in which $N$ is the number of e-folds of inflation \cite{Abolhasani:2013zya, Dulaney:2010sq,  Gumrukcuoglu:2010yc, Watanabe:2010fh,  Bartolo:2012sd, Emami:2013bk, Shiraishi2012,  Lyth:2013sha, Chen:2014eua}.  Imposing the recent Planck data \cite{Kim:2013gka}
(see also  \cite{Ramazanov:2013wea}) one obtains $| g_*| \lesssim 10^{-2}$.

Comparing the two power spectra Eqs. (\ref{quad-P}) and (\ref{dipole-power}) we stress that they have distinct structural forms. First, the quadrupole asymmetry Eq. (\ref{quad-P}) is in Fourier space while the dipole asymmetry  Eq. (\ref{dipole-power}) is in real space. Second, Eq. (\ref{quad-P}) represents anisotropy at each point on  CMB sphere so it does not distinguish between two different CMB hemispheres. However, Eq. (\ref{dipole-power}) represents asymmetry between two CMB hemispheres, i.e.  the asymmetry  depends on the direction of the look to sky given by the unit vector ${\bf \hat n}$.

Motivated by the quadrupole asymmetry obtained for the specific model of  anisotropic inflation \cite{Watanabe:2009ct} (see also \cite{Ackerman:2007nb} for a different realization of quadrupole anisotropy) we can consider the following general presentation of statistical anisotropies in power spectrum \cite{Pullen:2007tu}
\begin{equation}
\label{Eq-psaniso0}
\calP({\bf k}) = \overline{\calP_\calR } (k)   \left[1+\sum_ {LM}g_{LM}(k)Y_{LM}(\hat{\bf k})\right],
\end{equation}
where  $Y_{LM}(\hat{\bf k})$ (with $L\geq 2$) are spherical harmonics and the parameter $g_{LM}(k)$ quantifies the departure from statistical isotropy as a function of wavenumber $k$. Imposing the reality of the power spectrum one obtains
$g^*_{L M}=(-1)^L g _{L-M} $.  For our specific anisotropic inflation given by the action Eq. (\ref{action}) we have  $g_{2 0} \propto g_* $ while the rest of $g_{LM}$ are zero.

\section{Effects of anomalous inflationary features on the  abundance of LSS }
\label{Sec-III}


\begin{figure}
\includegraphics[scale=1.2]{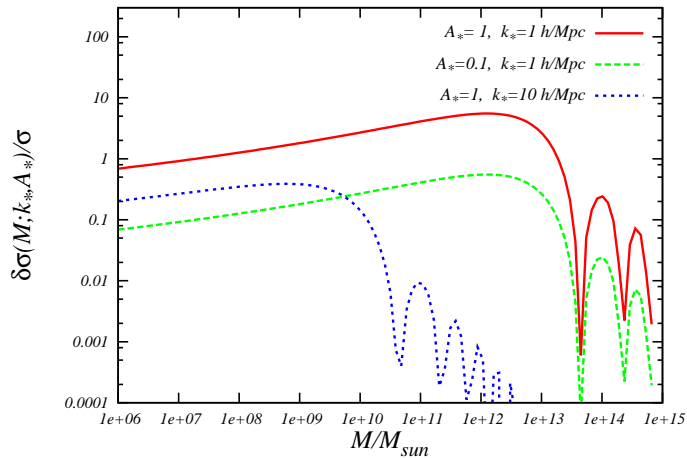}
\caption{ The ratio of change in variance to the variance in $\Lambda$CDM $, \delta \sigma/\bar \sigma$, induced from localized feature. The curves are  for localized feature  introduced in Eq. (\ref{Eq-delta}) with   $A_*=1 h/Mpc$ ,  $k_*=1 h/Mpc$ (red-solid curve);
$A_*=0.1 h/Mpc$,  $k_*=1 h/Mpc$ (green dashed curve) and $A_*=1 h/Mpc$ , $k_*=10 h/Mpc$ (blue dotted curve).  }
\label{fig:variance-ratio}
\end{figure}


Having studied the effects of the primordial features on the number density of structures   in Section \ref{Sec-II} and after presenting three different types of features in inflationary power spectrum in Section \ref{theory}, in this Section we study the effects of these  features on the variance of linear order perturbations and consequently on the number density of the structures. We study the predictions of each feature in turn.


\subsection{The Localized Feature}

The matter power spectrum of structures are nearly well constrained for the scales in the range $0.01 h Mpc^{-1}<k<1 h Mpcs^{-1} $ by large scale structure surveys like Sloan Digital Sky Survey (SDSS)-III  Baryon Oscillation
Spectroscopic Survey (BOSS) project\cite{Anderson:2013zyy}. However, beyond the mentioned scales, the data are not strong enough  to put constraints on the matter power spectrum. For $ k <0.01 hMpc^{-1}$ the volume of the sky to find the statistics is much bigger than the coverage of the ongoing surveys. On the other hand, for  scales $k>1 h Mpc^{-1}$, there is a lack of statistics for the dwarf satellite galaxies  to get a significant limit measurement for the power. Accordingly, constraining observationally the matter power spectrum on small scales  is a challenging task. There are many proposals to detect the dark matter sub-halos gravitationally which can be used to put constraints on the matter power spectrum on small scales \cite{Baghram:2011is,Rahvar:2013xya}. Also the detection/null detection of gamma rays that are anticipated from the annihilation of dark matter particles can constrain the matter power spectrum and the initial curvature perturbation power respectively \cite{Scott:2012kx}.

From the  theoretical point of view, as discussed in Sec.(\ref{Sec-I}), the primordial power spectrum of curvature perturbations can be translated to the number density of the structures in the Universe. Consequently, constraining the mass profile of the structures or statistics of dark matter tracers can be used as a probe of primordial curvature perturbation.

Now consider the local feature in power spectrum given by Eq. (\ref{Eq-delta}) localized at the scales
$k=k_*$ with the amplitude $A_*$ in units of $hMpc^{-1}$.  As mentioned before, $A_*$ can take either signs.  The change in variance from the general formula Eq. (\ref{Eq:variance-change})  is obtained to be
\be \label{Eq:deltasigma}
\delta\sigma =\frac{2A_*}{25\sigma}\frac{k_*^3T^2(k_*)D^2(z)}{\Omega_m^2 H_0^4}{\cal{P}}_{\cal{R}}(k_*)W^2(k_*R)
\ee
 In Fig. \ref{fig:variance-ratio}, we have plotted the ratio of the change in variance  to variance in $\Lambda$CDM  model, ($\delta\sigma/\bar{\sigma}$), for  this feature.  The red solid curve, the green dashed curve and the blue dotted curve respectively  are for  $A_*=1 h Mpc^{-1}$  at $k_*=1 h/Mpc$, $A_*=0.1 hMpc^{-1}$ at $k_*=1 h/Mpc$ and $A_*=0.1 hMpc^{-1}$ with $k_*=10 h/Mpc$.
 It is obvious that  $\delta \sigma$ linearly depends on the amplitude of $\delta {\cal{P}_\calR}, A_*$.
 It is worth to mention that the oscillatory features in Fig. \ref{fig:variance-ratio}  for high masses are not physical. These oscillatory features are caused by the choice of the window function in Eq.(\ref{Eq:deltasigma}). Accordingly the physical feature is for the mass range $M<M_c$, where  $k_*\simeq 1/R_*=[4\pi\bar{\rho}/3M_c]^{1/3}$.   On the other hand,  the position of feature affects the amplitude of  $\delta \sigma$ too.

The window function smoothes all the modes smaller from the specific mode $k_*$ in the Gaussian window function in Fourier space. For each specific mode $k_*$, we can find the corresponding critical mass $M_c$ as follows
\be
M_c=\frac{H_0^2}{2G}(k_*)^{-3}\Omega_m(z)
\ee
where $H_0$ is the Hubble constant and $\Omega_m$ is the density parameter of matter in redshift $z$ given by
\be
M_c=1.1*10^{12} h^2 (\frac{k_*}{h / Mpc})^{-3} \Omega_m(z) \, .
\ee
For an inflationary model with $k_*=1 h/Mpc$, the effect of feature on the number count of the structures would be valid for    $M_c (z) < 1.5 \times 10^{11}(1+z)^3$, where for the redshift $z<1$ the critical mass is less than the Milky way mass. Accordingly,  the number density of Milky way type galaxies can not be used as a viable observational constraint on the amplitude of the localized feature  for the wavenumber $k_*\geq1 h/Mpcs$.

 If the feature is placed in smaller wavenumbers (larger scales) it
affects $\delta \sigma$ more significantly. This is because the matter transfer function decays like $\approx\ln k/k^2$ so it obtains less contributions from power spectrum modifications on small scales.
Another  important point to notice is that  $\delta\sigma/\sigma$ is independent of redshift. As a result the redshift-dependence of the modifications in abundance comes from the universality function and not from the change in variance.

In the next step, we calculate the change in number density, $\delta_n$, for both Press-Schechter and
the Sheth-Tormen mass functions. For the  Press-Schechter mass function we obtain
\be
\delta_n |_{PS}(M,z;A_*,k_*) = \left(\nu^2(M,z)-1\right)\frac{\nu^2(M,z)}{\delta^2_c}\frac{2 }{25}\frac{A_*k_*^3T^2(k_*)D^2(z)}{\Omega_m^2 H_0^4}{\cal{P}}_{\cal{R}}(k_*,z)W^2(k_*R) \, ,
\ee
while for  for the Sheth-Tormen universality function the result is
\be
\delta_n |_{ST}(M,z;A_*,k_*) = \left[a\nu^2(M,z)-1+\frac{2ap\nu^2(M,z)(a\nu^2(M,z))^{-p-1}}{[1+(a\nu^2(M,z))^{-p}]} \right] \frac{2\nu^2(M,z) A_*k_*^3T^2(k_*)D^2(z)}{25\delta^2_c \Omega_m^2 H_0^4}{\cal{P}}_{\cal{R}}(k_*,z)W^2(k_*R)
\ee

\begin{figure}
\includegraphics[scale=1.2]{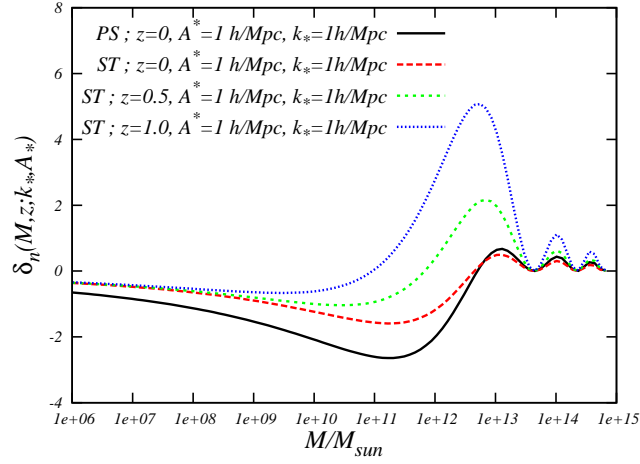}
\caption{The change in number density versus the mass of structures
from the  feature localized at $k_*=1 h/Mpc$ with the amplitude $A_*=1 h/Mpc$ for Press-Schechter mass function at redshift $z=0$ (black solid line) and  for Sheth-Tormen mass function at $z=0$, $z=0.5$ and $z=1$ shown respectively by the  red long dashed curve, the green dashed curve and the blue dotted curve.}
 \label{fig-n-ratio}
\end{figure}

The change in number density, $\delta_n$, from the localized feature is plotted in Fig. \ref{fig-n-ratio}. In this figure we plot $\delta_n$ for Press-Schechter and Sheth-Tormen universality functions versus the mass scale. The feature is located at $k_*=1 h/Mpc$ with the amplitude of $A_*=1 h/Mpc$. The  black solid (red long dashed) curve is for Press-Schechter (Sheth-Tormen)
universality function at $z=0$. These two curves show that the change in abundance is smaller for the more realistic $f(\nu)$ function in Sheth-Tormen proposal. Therefore, in the follow up figures we present  the Sheth-Tormen prediction for $\delta_n$.
The Green-dashed (blue dotted) curve in Fig. \ref{fig-n-ratio} is the prediction for $z=0.5$ ($z=1.0$).   Fig. \ref{fig-n-ratio}  shows that in the mass range of $M<M_c$ where the variance change is physical and not the artifact of window function, $\delta_n$ decrease with redshifts. This means that we  find the
most noticeable deviation from the standard case at smaller redshifts (present time). This can be explained by the fact that the variance becomes larger at small redshifts so the height parameter decreases. Accordingly,  as discussed in Sec.(\ref{Sec-II}), for the range $M<M_c$ ( $\nu<\nu_c$)   we have more deviation from standard case.   Note that, as in the case of $\delta \sigma$, the oscillatory patterns for higher mass ranges are the artifact of the window function.

As a result of  Eq. (32) the change in the number density of the structures, $\delta_n$, is proportional to $(\nu^2-1)\, A_*$.
 This means that for the mass range with $\nu>1$ the positive feature in the power-spectrum increases the number of structures in comparison to the standard case, while for the mass range $\nu<1$ we have the opposite effect. The main point here is that the regime $\nu<1$ corresponds to the collapse regime where $\sigma(M)>\delta_c$ and the actual structures are formed. This is the regime where the positive bump in primordial power-spectrum reduces the number of structures. The intuition for this comes from the probability distribution function of primordial density seeds (like in Press-Schechter case $P_{\cal{R}}({\delta})=\frac{1}{\sqrt{2\pi}\sigma_R}e^{-\delta^2/(2\sigma_R^2)}$), where the probability is maximum for a fixed density contrast when $\delta=\sigma$. In other words the first crossing rate in PS formalism depends on $\nu\times\exp(-\nu^2)$ and  accordingly the change in variance has an extremum  at $\nu=1$ as we anticipated from the above discussions.


\begin{figure}
\includegraphics[scale=1.2]{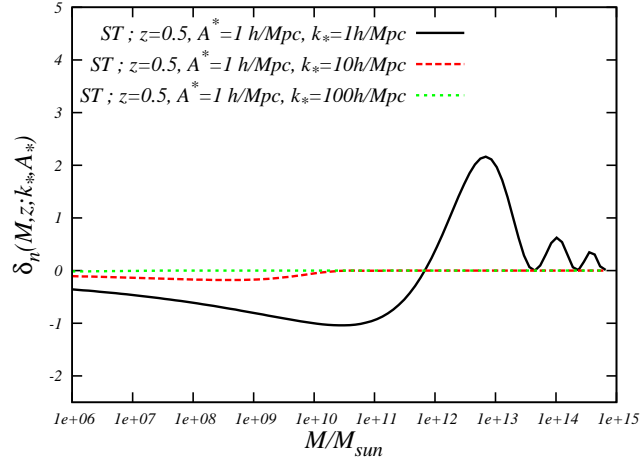}
\caption{The change in number density from the localized  feature versus the mass of structures at redshift $z=0.5$ for Sheth-Tormen universality mass function. The features have  the amplitude $A_*=1 h/Mpc$ with $k_*=1 h/Mpc$ (solid black curve);   $k_*=10 h/Mpc$ (red long dashed curve);  $k=100 h/Mpc$ (green dashed curve). }
\label{fig-n-ratioA}
\end{figure}


In Fig. \ref{fig-n-ratioA}   we have plotted the density contrast versus mass at redshift $z=0.5$ for features localized  at  $k_*=1 h/Mpc$, $k_*=10 h/Mpc$ and $k_*=100 h/Mpc$. It is obvious that by changing the position of feature  we get different mass scales $M_c$ for the  deficiency in the number of structures for a given value of  $A_*$. Also the amplitude of $\delta_n$ decreases by  increasing $k_*$.  An important  point to mention is that the matter power spectrum is nearly well constrained for the scales in the range   $0.2 hMpc^{-1}\le k\le 1 h Mpc^{-1}$ by the Lyman-alpha forest  \cite{Croft:2000hs,Gnedin:2001wg,Tegmark:2002cy,Seljak:2004xh}. Accordingly, there are flexibilities in  turning  on features on smaller scales with $k> 1 h Mpc^{-1}$. In Figs.  \ref{fig-n-ratio} and \ref{fig-n-ratioA}, we observe that a feature with the positive amplitude, $A_* >0$, reduces the number density of structures for the mass  smaller  than $M_c\sim \bar{\rho}k^{-3}_*$.
As a result, the number of structures decreases  by the amount
\be
\delta N(M_i<M <M_f)= f_{sky}\int _0^{z_{obs}} dz \int _{M_i}^{M_f} \bar{n}(M,z)\delta_n (M,z;A_*,k_*)\frac{dV}{dz}
\ee
where $\delta N= N( A_*)-\bar{N}(A_*=0)$ is the change in the number of the structures,  $f_{sky}$ is the fraction of the sky in which a specific survey collects the data and   $M_{i} \,  (M_f)$ is the initial (final) mass range  which we are interested in. The volume integration is substituted with integral over redshift as the cosmological volumes are redshift-dependent ($dV=\frac{\partial V}{\partial z}dz$).

This mechanism provides a chance to resolve  the missing satellite problem, the  tension which is raised between the simulations and observations on the number density of dwarf galaxies \cite{ Klypin:1999uc,Kravtsov:2003sg,Bullock:2010uy}. In order to address this discrepancy, different proposals were put forward. One proposal is to look for astrophysical solution, asserting that there is a minimum dark matter halo mass below which halos can not host a  galaxy. This is because the gravitational potential of the dark matter halo is not sufficient to encapsulate the baryons in it. Astrophysical processes like SNIe explosion and AGN feedbacks wash the baryons out of the gravitational potential of matter   \cite{Brooks:2012ah}.  The second proposal is assuming that dark matter halos themselves do not exist in these scales. This can be explained if  dark matter is not cold or not completely dark such as in  warm dark matter and self-interacting dark matter models  \cite{Goetz:2002vm,Vogelsberger:2014pda}.
It is worth to mention that there are proposals to detect the dark matter substructures which are not hosting baryonic matter with gravitational lensing technique \cite{Vegetti:2009cz}.  Our model based  on local feature in inflationary power spectrum can be considered as the third proposal.  In our proposal, by changing  the variance of  dark matter perturbation, we can change the statistics of the  structures in the Universe. This proposal requires further cosmological observations to be tested.


\subsection{Primordial Dipole Asymmetry}

In this subsection we study the effects of dipole asymmetry in primordial power spectrum on the statistics of the structures. As discussed in subsection \ref{dipole-sec} there are indications of dipole asymmetry in CMB map as observed by Planck collaboration \cite{Ade:2013nlj}. The primordial power spectrum with dipole modulation is given in Eq. (\ref{dipole-power}).

First we consider the simple case in which the amplitude of dipole asymmetry, $A_d$, does not depend on scales. In this approximation  the change in the
variance from the general formula Eq. (\ref{Eq:variance-change}) is obtained to be
\be
\delta\sigma(M;A_d,\hat{\bf p})=A_d\cos(\theta_{\hat{\bf n}.\hat{\bf p}}){\bar{\sigma}(M)} \, .
\ee
This indicates  that $\delta\sigma$ depends on both the amplitude of the dipole and the direction of the observation $\hat{\bf n}$. The maximum (minimum) change in variance occurs along the direction of dipole (perpendicular to direction of dipole).

Now we can  find the change in the statistics of the structures.  For the Press-Schechter  mass function
the change in number density is obtained to be
\be
\delta_n |_{PS}(M; A_d,\hat{\bf p})  = \left(\nu^2(M)-1 \right)A_d \cos(\theta_{\hat{\bf n}.\hat{\bf p}}) \, ,
\ee
while  for the Sheth-Tormen case we have
\be
\delta_n |_{ST}(M; A_d,\hat{\bf p}) = \left(a\nu^2(M,z)-1+\frac{2ap\nu^2(M,z)(a\nu^2(M,z))^{-p-1}}{[1+(a\nu^2(M,z))^{-p}]}   \right)A_d  \cos(\theta_{\hat{\bf n}.\hat{\bf p}}) \, .
\ee

Below we present the predictions for $\delta_n$ with the dipole amplitude  $A_d=0.07$ and $\hat{\bf p}=(l=227,b=-27)$ in galactic coordinate as reported by the Planck observation \cite{Ade:2013nlj}.

\begin{figure}
\includegraphics[scale=1.2]{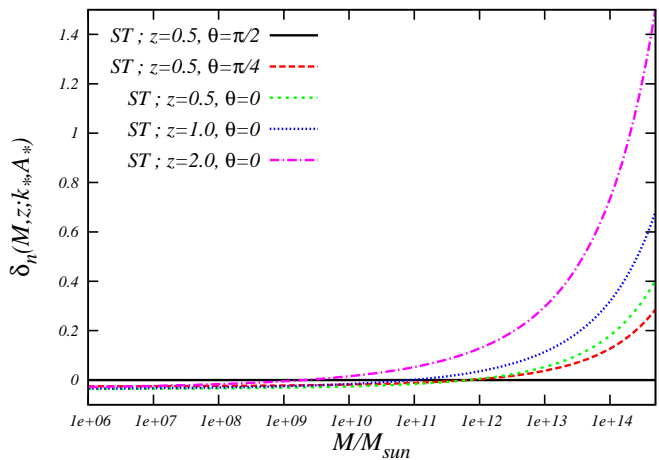}
\caption{$\delta_n$ vs. mass for the model with dipole asymmetry.   The black solid line and the  red long dashed curve are for $\theta_{\hat{n}.\hat{p}}=\pi/2$ and $\theta_{\hat{n}.\hat{p}}=\pi/4$ at the redshift $z=0.5$ respectively. The predictions for $\theta_{\hat{n}.\hat{p}}=0$ at redshifts $z=0.5$, $z=1.0$ and $z=2.0$ are presented by green dashed, blue dotted and magenta dash-dotted curves respectively.}
\label{fig-n-ratio-dipole}
\end{figure}


\begin{figure}
\includegraphics[scale=1.2]{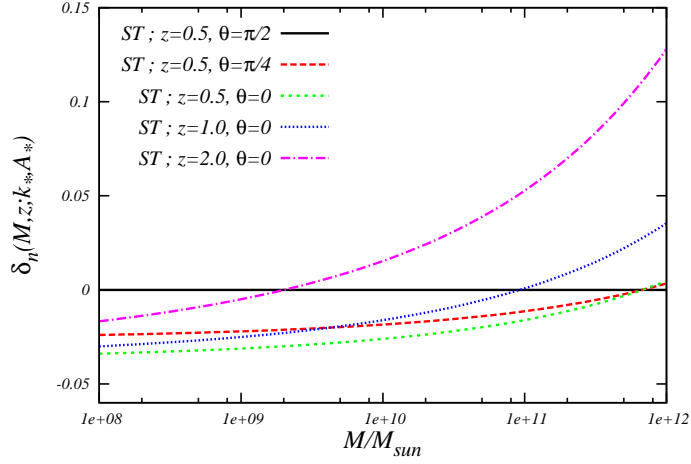}
\caption{ This Figure is the zoom in of Fig. \ref{fig-n-ratio-dipole} with the same caption.}\label{fig-n-ratio-dipole1}
\end{figure}


In Fig. \ref{fig-n-ratio-dipole} we plot $\delta_n$ versus mass.  The horizontal black solid line represents   the prediction at redshift $z=0.5$ with the observation direction angle $\theta_{\hat{n}.\hat{p}}=\pi/2$. This curve shows that for the direction of sight perpendicular to dipole direction  the number density of the structures does not change compared to the standard case. The red long dashed
and the green dashed curves respectively are for  $\theta_{\hat{n}.\hat{p}}=\pi/4$ and $\theta_{\hat{n}.\hat{p}}=0$ at the redshift $z=0.5$. The blue dotted and the magenta dash-dotted curves respectively are for redshifts $z=1$ and $z=2$ at angle $\theta_{\hat{n}.\hat{p}}=0$.  The essential point here is that $\delta_n$ increases with redshift while it also depends on $\theta_{\hat{n}.\hat{p}}$,  the angle between the direction of the observation and the dipole direction in sky.

In Fig.  \ref{fig-n-ratio-dipole1} we zoom in mass scale where one can see the $\nu_c$ barrier crossing at different redshifts. The critical mass, where $\delta \sigma$ changes sign, increases with redshift. This is understandable intuitively  because of the hierarchy in structure formation. When one is in the region $\nu<\nu_c$ at higher redshifts one gets less deviation from standard case. This is because at higher redshifts the growth function is smaller and consequently the power spectrum and the variance are smaller. This means that the height parameter approaches the  critical value and the effects of dipole corrections become small. On the other hand, when one crosses the $\nu_c$  barrier again in higher redshift then the height parameter becomes larger but it deviates from $\nu_c$ and  consequently one obtains more structures.

As discussed in subsection \ref{dipole-sec} the CMB dipole asymmetry shows non-trivial scale-dependence such that for scales $\ell > 600$ there is the upper bound on the amplitude of dipole
$A_d < 0.017$ from the quasar data \cite{Hirata:2009ar}. Therefore, it is desirable to consider the predictions of a scale-dependent dipole asymmetry on the abundance of structures.  The simplest proposal is to assume a  power-law behavior for $A_d(k)$ as follows
\be \label{Eq:powerdiprun}
{\cal{P}}_{{\cal{R}}}=\overline{\calP_\calR} \left[1+2A_{dp}\left(\frac{k}{k_{dp}} \right)^{n_A-1}\cos(\theta_{\hat{\bf n}.\hat{\bf p}}) \right] \, ,
\ee
where $A_{dp}$ is the amplitude of the dipole asymmetry at the pivot wavenumber $k_{dp}$ and $n_A$ shows the spectral index of the dipole amplitude. For $n_A=1$ we obtain the
scale-invariant dipole asymmetry studied above.
Now substituting the primordial power spectrum Eq. (\ref{Eq:powerdiprun}) in Eq. (\ref{Eq:variance-change}),  $\delta \sigma $ is obtained to be
\be
\delta\sigma(M; A_d,\hat{\bf p})=A_{dp}\cos(\theta_{\hat{n}.\hat{p}})/(\bar{\sigma}(M))\int{dk}k^{n_A+2} \frac{4}{25}\frac{T^2(k)D^2(z)}{k^{n_A-1}_{dp}\Omega_m^2 H_0^4}{\cal{P}}_{\cal{R}}(k)W^2(k) \, .
\ee
Correspondingly,  the change in number density is given by
\be
\delta_n(M; A_d,\hat{\bf p}) = \left(\nu^2(M)-1 \right)\frac{\nu^2}{\delta^2_c}A_{dp}\cos(\theta_{\hat{n}.\hat{p}})\int{dk}k^{n_A+2} \frac{4}{25}\frac{T^2(k)D^2(z)}{k^{n_A-1}_{dp}\Omega_m^2 H_0^4}{\cal{P}}_{\cal{R}}(k)W^2(k) \, .
\ee


\begin{figure}
\includegraphics[scale=1.2]{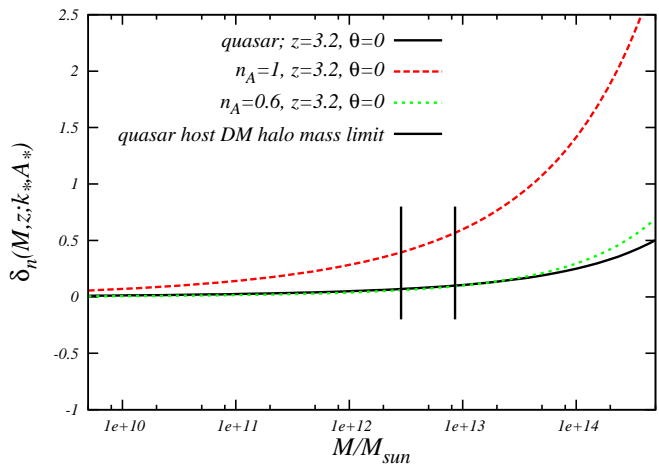}
\caption{ $\delta_n$ vs. mass for the model with dipole asymmetry at $z=3.2$. The red long dashed curve and the green dashed curve  show the prediction for   $n_A=1$ and  $n_A=0.6$ with $\theta_{\hat{n}.\hat{p}}=0$ respectively. The black solid curve shows the constraints from quasars on dipole asymmetry. The vertical solid black lines indicate the mass ranges of dark matter halos which are the hosts of quasars. The mass-range of quasar host halos are $M=(2-6) \times 10^{12} h^{-1} M_{\odot}$. }
\label{fig:running}
\end{figure}

 Hirata \cite{Hirata:2009ar} has used the quasar data to check the amplitude of hemispherical asymmetry in LSS in which  the constraint $\nabla\sigma_8/\sigma_8\sim 0.027 r_{lss}^{-1}$ at redshift  $z=3.2$ is interpreted as the  bound  $A_d<0.017$ with 1-sigma accuracy marginalized over directions. Now we can use this  observation to constrain the spectral index of the dipole asymmetry. In Fig. \ref{fig:running}  we have plotted $\delta_n$ versus mass for Sheth-Tormen universality function at redshift $z=3.2$. The solid black curve represents the limit obtained from quasar analysis. In order to constrain the spectral index of dipole anisotropy, we should consider the mass between
 $M$ and  $M+dM$ of the host halos of quasars. The minimum mass of the host dark matter halo typically are $M\simeq (2-6)\times 10^{12}h^{-1} M_{\odot}$ \cite{Shen2007}. As a  result  we obtain the bound  $n_A<0.6$ to satisfy the LSS constraints on the dipole asymmetry.
 It is important to mention that we assume  the quasars are a fair sample of galaxies with a host halo of masses $(2-6) \times 10^{12} h^{-1} M_{\odot}$. Accordingly counting the number of quasar in a specific redshift bin
gives us  information on the distribution of dark matter halos. Then we assume that the amplitude of  dipole asymmetry seen in quasar data, $A<0.017$,  can be assigned as an upper limit for  $A$ obtained from the  number count of dark matter halos. We change the spectral index of dipole asymmetry in the primordial power spectrum in order to have exactly the same amount of structures predicted by the observed upper limit dipole asymmetry in quasar data.
The value of $n_A < 0.6$ is the upper bound for dipole spectral index in that sense.
The mass range that we can trust our bound is  $M_c (z) < 1.5 \times 10^{11}(1+z)^3$. The quasar sample is in redshift $3.2$ ($M_c\simeq 10^{13}$), accordingly the mass range of the host dark matter halos are in valid range of our calculations.

\subsection{Statistical Anisotropies}

In this subsection we consider models of inflation which generate statistical anisotropies in curvature perturbation power spectrum  as given in Eq. (\ref{Eq-psaniso0}).  The case with $ L=2,M=0 $ is the known quadratic anisotropy \cite{Ackerman:2007nb} which is generated in models of anisotropic inflation \cite{Watanabe:2009ct}.

As in previous examples,  to calculate the change in  the abundance of structures,  first we calculate the change in the variance. The important point in this type of feature is that the variance is obtained from the integration in $k$-space so we can use the orthogonality of spherical harmonics to obtain \cite{Baghram:2013lxa}
\be
\delta\sigma= \frac{g_{00}}{2\sqrt{4\pi}}\bar{\sigma}(M) \, .
\ee
This means that in anisotropic models only the amplitude of the  variance is modified.

For  Press-Schechter mass function  $\delta_n$ becomes
\be
\delta_n |_{PS}(M; g_{LM})  = \left(\nu^2(M)-1 \right)\frac{g_{00}}{2\sqrt{4\pi}} \, ,
\ee
while  for the Sheth-Tormen mass function  we have
\be \label{Eq:deltanqu}
\delta_n(M; g_{LM}) |_{ST} = \left(a\nu^2(M,z)-1+\frac{2a\, p\, \nu^2(M,z)(a\nu^2(M,z))^{-p-1}}{[1+(a\nu^2(M,z))^{-p}]}   \right)\frac{g_{00}}{2\sqrt{4\pi}} \, .
\ee
Eq.(\ref{Eq:deltanqu}) shows that although the variance has changed by a constant amplitude ${g_{00}}/{2\sqrt{4\pi}}$ but the abundance of structures does not change just by a constant amplitude in which it  depends non-trivially  on the mass of structures which appears implicitly in the height parameter in Eq. (\ref{Eq:deltanqu}). As in previous examples, define
$\nu_c$ the point  where $\delta \sigma$ changes sign with the mass $M_c$. For the mass scales $M<M_c$ with  $\nu<\nu_c$ a positive (negative) change in the amplitude of the power-spectrum $g_{00}$ causes a decrease (increase)  in the abundance of  structures.
Interestingly, we see that quadrupole asymmetry with $L=2, M=0$ does not affect the abundance
of structures.

Although the anisotropic power spectrum introduced in Eq. (\ref{Eq-psaniso0}) does not change the statistics of the structures but it has a secondary effect through the bias parameter \cite{Baghram:2013lxa}. The bias parameter relates the abundance of dark matter halos, which are the host of galaxies (observable quantities), to the underlying dark matter density perturbations. Therefore,
models with anisotropy in the form of Eq. (\ref{Eq-psaniso0})  can have non-trivial predictions for  galaxy number density due to bias and not from the statistics of the structure such as in  Press-Schechter formalism.


\section{Conclusion and Discussions}
\label{Sec-IV}

In this work we studied the effects of  inflationary models with anomalous feature in
power spectrum  on the number density of  large scale structure and  dark matter halos which are  the host environment of galaxies and group of galaxies. The mass function of  structures in Universe is a function of local collapsed parameter (i.e. $\delta_c$) and the variance of mass distribution. The latter itself is the Fourier transform of matter density contrast power spectrum  which is a function of  the mass scale.  Therefore, the variance is related to the underlying cosmological model and the initial conditions. We have shown that any feature in primordial power spectrum induces a change in the variance of structures in large scales. Accordingly,  in the light of   Press-Schechter idea (to obtain the number density of structures), a change in variance results in a
change in number density of structures. We also assumed that to leading order the features in primordial power spectrum only modify the variance and the processes of Excursion Set Theory do not change  compared to standard case as we discussed at the end of  in Sec. \ref{Sec-II}.

First we have shown that the change in number density, $\delta_n$, is proportional to the change in variance, $\delta\sigma/\sigma$. However the sign of this relation depends on the mass scale and the redshift which one observes the structure. At any specific redshift, there is a mass scale $M_c$ in which  for   $M>M_c (M< M_c) $ corresponding to $\nu > \nu_c (\nu < \nu_c)$ we have  $\delta_ n\propto + \delta\sigma/\sigma$ $ (\delta n\propto - \delta\sigma/\sigma )$.  It is important to note that the regime  $\nu > \nu_c$  corresponds to a quasi linear regime, where we should interpret the results in this regime with cautious and the results in this work does not apply for $\nu\gg\nu_c$ as the structures are not yet formed at these scales.

Three distinct anomalous features are considered in Sec. \ref{Sec-III}. First we considered models with localized feature in power spectrum. We have shown that  localized features with positive amplitude (excess in power spectrum)  can reduce the number of the structures at corresponding scales. While the power spectrum of matter distribution is well constrained at scales $k<1 h Mpc^{-1}$  but  the missing satellite problem at galactic scales opens up the possibility that the  inflationary models with features can be a promising new candidate to solve the cold dark matter challenges on galactic scales. The other interesting point is that the statistics of structures decreases with redshift. Consequently, we have the maximum effect from the localized feature in local Universe.

Next we have investigated the effects of primordial curvature perturbation with dipole asymmetry on the abundance of structures. The change in the abundance of structures is proportional to the amplitude of the dipole asymmetry. There is no change for the abundance of structures in the direction perpendicular to the dipole direction while this change is maximum along the direction of dipole.  In the limit  $\nu<\nu_c$ the change in abundance is larger at low redshifts while for  $\nu>\nu_c$ this behavior is reversed.
In order to be consistent with observations  on sub-CMB scales we have considered dipole asymmetry
in the form  $A=A_{dp}(k/k_{dp})^{n_A-1}$. Imposing the constraints from quasar data we have found $n_A<0.6$. Finally we studied  the imprints of
primordial curvature perturbations with statistical anisotropies, such as in models of anisotropic inflation, on the abundance of structures.  We have shown that the change in the amplitude of variance is controlled by $g_{00}$ while the change in abundance of structures depends only on the universal mass function.

The large scale structure observations are promising arena for observational cosmology to probe the physics of early universe. Any specific deviation from the standard  nearly adiabatic, nearly
scale-invariant, nearly Gaussian and isotropic   initial conditions can be probed in LSS. In addition  the  scale-dependence of these features (i.e. Non-Gaussianity, dipole asymmetry etc) can be probed by the tracers of dark matter. However, non-linear structure formation, the mechanism of collapse, dark matter and baryon biases and redshift space distortions all are the complications which need to be understood in order to investigate the physics of early universe.
\vspace{1cm}

{\bf {Acknowledgment}} \\

We would like to thank  Razieh Emami and Nima Khosravi for useful discussions and comments.

\end{document}